\newif\ifPDF
\ifx\ifPDF\undefined

  \documentclass[acmtoce]{acmtrans2m}

  \usepackage[dvips]{graphicx}
\else
  \documentclass[acmtoce]{acmtrans2m}
  \usepackage[pdftex]{graphicx}
  \DeclareGraphicsExtensions{.pdf,.png,.jpg}
\fi
\pdfoutput=1

\usepackage{graphicx}
\graphicspath{{./img/}}
\usepackage{amsmath}
\usepackage{array}
\usepackage{enumerate}
\usepackage{lscape}
\usepackage{placeins}

\usepackage{subfigure} 

\newcommand{\sys}[1]{\textsc{#1}\index{\textsc{#1}}}
\newcommand{\ac}{\sys{AC}}

\title{Uncovering Plagiarism Networks}

\date{}

\author{MANUEL FREIRE, MANUEL CEBRI\'AN and EMILIO DEL ROSAL\\Universidad Aut\'onoma de Madrid}

\begin{abstract}
Plagiarism detection in educational programming assignments is still a problematic issue in terms of resource waste, ethical controversy, legal risks, and technical complexity. This paper presents \ac{}, a modular plagiarism detection system. The design is portable across platforms and assignment formats and provides easy extraction into the internal assignment representation. Multiple similarity measures have been incorporated, both existing and newly-developed. Statistical analysis and several graphical visualizations aid in the interpretation of analysis results. The system has been evaluated with a survey that encompasses several academic semesters of use at the authors' institution.
\end{abstract}







\begin{document}

\setcounter{page}{111}

\begin{bottomstuff}
Author's address: Escuela Polit\'ecnica Superior,
Universidad Aut\'onoma de Madrid,\newline
28049 Madrid, Spain.
\texttt{\{manuel.freire, manuel.cebrian, emilio.delrosal\}@uam.es}
\newline
\end{bottomstuff}
\maketitle

\section{Introduction}
\label{intro}

The development of the World Wide Web and the associated abundance of accessible electronic documents has lead to a greater incidence of plagiarism in undergraduate studies. Exact figures are unknown, since successful plagiarism is by definition not detected, but are believed to be high and growing \cite{clare2000cpt,irving2000pde}. Alexander Aitken, one of the leading experts in operating plagiarism detection software, asserted in a personal communication prior to 2001 that for any (USA) student corpus, 10\% of submissions are plagiarized \cite[p.~4]{culwin2001scp}. Manual plagiarism detection is tedious and extremely time consuming for educators. Additionally, it is emotionally and legally risky for student and educator alike \cite{harris1994pcs}. University experience also shows that even minor plagiarism levels can cause a mistrust for the work of students which can lead to baroque examinations to prove the authenticity of each student's work, or to the relative weight of possibly-plagiarized submissions in the final grade to be far lower than their actual share of effort would warrant. Ignoring the problem posed by plagiarism results in unfair grading, and can have an avalanche effect in plagiarism incidence levels.

Plagiarism detection and prevention can be examined from two points of view. The first is of ethical and normative nature, and involves addressing the deeper causes of plagiarism and selecting appropriate academic and legal deterrents. This perspective is examined, for example, in \cite{Braumoeller} and \cite{martin1994pme}. The second perspective emphasizes the technical measures required to actually detect plagiarism. Both perspectives are complementary, because heavy penalties require non-negligible chances of detection to be effective.

Detecting plagiarism requires the comparison of large collections of documents to each other. Automated plagiarism detection systems can perform these comparisons in a matter of seconds, allowing graders to concentrate on the most likely cases of plagiarism. For the same amount of effort, a grader using automated tools can achieve a much higher detection ratio. This results in a strong deterrent for potentially dishonest students.

To reliably detect plagiarism, automated detection systems need to have access to both the original and plagiarized submission. This presents a difficult problem in non-software plagiarism, where systems are forced to build huge document corpora from whatever online sources students may have access to. However, in the case of of software plagiarism, ``paper mills'' found in other disciplines do not exist, because each software assignment is expected to work under very different conditions, and integrating code from external sources into a working application is generally more demanding than original work. The present work focuses exclusively on the problem of source-code plagiarism detection.

This paper presents \ac{}\footnote{available online at \textbf{http://tangow.ii.uam.es/ac}}(\textsc{Anti-Copias}), a source-code plagiarism detection tool. \ac{} offers many improvements over other tools described in current literature: the use of rich visualization greatly simplifies the task of analyzing the result of similarity tests; its stand-alone implementation does not raise privacy concerns found in web-based systems; and preparation of assignment submissions for plagiarism detection, overlooked by many systems, is partially automated with a graphical user interface.
The remainder of this paper is organized as follows: Section \ref{state} motivates the development of \ac{} by means of a review of the current state of the art in software plagiarism detection; Section \ref{design} briefly describes \ac{}'s design. Section \ref{eval} presents an evaluation of \ac{}, both formal and informal. Finally, Section \ref{conclusion} presents conclusions and outlines future work.

\section{Motivation and State of the Art}
\label{state}


No grader can be reasonably expected to perform the 5050 possible pairwise comparisons required to check for plagiarism within a corpus of 100 submissions; a plagiarism detection system running on a standard desktop computer, on the other hand, can perform such a check in seconds. However, computers cannot assign intent to their findings. Two submissions may be similar for perfectly legit reasons (for instance, the use of instructor-supplied code). The value of automated plagiarism detection lies in narrowing down the search, helping a grader concentrate on the most probable cases of plagiarism. Increased probability of detection significantly alters the cost-benefit analysis of a would-be plagiarizer. Quoting Braumoeller, \cite{Braumoeller}:

\begin{quote}
Warning students not to plagiarize, even in the strongest terms, appears not to have had any effect whatsoever. Revealing the use of plagiarism-detection software to the students prior to completion of an assignment, on the other hand, proved to be a remarkably strong (though still not absolutely perfect) deterrent.
\end{quote}

Even though plagiarism prevention can be seen as a matter of due diligence, most educators prefer to invest their time teaching rather than policing or investigating students. After locating an instance of plagiarism, a grader is typically expected to perform further inquiries, speak to the suspects, and apply the corresponding penalties. Frequently, cases of plagiarism are not completely clear-cut, and the finder is comfronted with the choice of applying too harsh a penalty or no penalty at all, a decision that most graders would prefer to avoid. The temptation to skip plagiarism detection is stronger when detection requires substantial effort. Graders that undertake this effort may be considered over-zealous, and, in addition, will encounter the moral hazard of having to apply penalties to students that would have safely avoided detection if graded by their peers. Manual plagiarism detection is an ungrateful chore.

Automated plagiarism detection can partially address this moral hazard. It can vastly simplify the effort required to locate cases of plagiarism, and provide additional backing evidence which would be difficult to acquire otherwise. For instance, an automated tool can easily locate the set of constructs that are unique to a given pair of submissions, and compare this against the average number of unique constructs for all other pairs the corpus that is being examined. Additionally, a tool can perform these checks in an objective manner, analyzing all submissions in its corpus without any bias or preconception. In this sense, the results will be more ``fair'' than those that would be achieved by comparing only those submissions which happen to catch a grader's fancy and discarding the rest. Finally, policies regarding the use of detection systems can be standardized across all graders of a particular course or set of courses, allowing all submissions, regardless of grader, to receive the same level of plagiarism-related scrutiny. In spite of these advantages, a 2001 survey of UK computing departments reported that only 26\% of respondents used automated plagiarism detection \cite{culwin2001scp}.


\subsection{State of the Art}

Several plagiarism detection systems have been implemented since the 1960s. \sys{MOSS} \cite{MOSS}, \textsc{JPlag} \cite{prechelt2002fpa}, \sys{SIM} \cite{SIM}, \sys{YAP} \cite{YAP} and \sys{SID} \cite{SID} are probably  the most widespread within the academic community. Lancaster and Culwin \cite{lancaster2004csc} enumerate a set of comparative properties which can aid a grader or institution in the selection of a plagiarism detection system. A modified version follows:

\begin{itemize}

\item \emph{Availability} -- many systems are no longer available. Those that are still in use may be private (restricted to their host institutions), or allow distribution to outside parties. Distribution may be only in binary format, or may include access to the actual source code.

\item \emph{Locality} -- systems may be remote or local. Remote systems are mostly web-based, and may include a client-side application to upload submissions. Local systems, on the other hand, may require a very specific environment to run.

\item \emph{Documentation and support} -- existence of technical information in the form of papers and reports, or even source code and associated program documentation; and degree of support that can be expected from the developers or current maintainers in the event of problems. Many systems are unmaintained.

\item \emph{Preprocessing} -- The inclusion of tools or interfaces that allow users to assemble the submission corpus into the format expected by the system.

\item \emph{Visualization} -- The quality of the interfaces used to present similarity results to users and allow analysis-related queries.

\item \emph{Algorithms and breadth} -- quality and breadth of the similarity computation algorithms incorporated into the tool. Additionally, certain algorithms require support for each particular programming language where they are to be used.

\end{itemize}

System \emph{locality} and \emph{availability} are gaining importance due to increased awareness of intellectual property issues. Data protection laws from several countries and universities limit the export of identifiable data to external organizations. This prevents graders from using external, web-based engines, unless additional measures are taken to anonymize submissions or the system is readily available to be installed on-site. \sys{MOSS} and \sys{JPlag} are both web-based and private. On the other hand, local systems are frequently geared towards highly specific environments, and many of them are either out of availability (\sys{TeamHandIn} \cite{culwin1995pap}, \sys{Saxon} \cite{saxon2000cpd}, \sys{Cogger} \cite{cunningham1993uct}), heavily platform-dependent, and/or unsupported (\sys{YAP3}, \sys{SIM}, \textsc{Sherlock} \cite{joy1999ppa}, \sys{Jones} \cite{jones2001mbp}, \sys{Big Brother} \cite{irving2000pde}).

Technical \emph{documentation} should fully describe a system, allowing effective use by graders. The effectiveness of many similarity detection algorithms is strongly dependent on implementation details such as thresholds and parsing methods. Some authors argument that cheaters could use detailed documentation to avoid detection, and therefore oppose public access to this information. This is akin to ``security by obscurity'', and prevents assessment of the system's true strengths and weaknesses. Additionally, should students within an institution learn that certain types of cheating are likely to be caught, while others seem to be much safer, the distribution of source-code manipulations in plagiarized programs will vary to reflect this perception. Updating the system to stop the leak requires the system to be actively maintained and well-supported, or empowering users to implement the necessary modifications themselves. This makes a strong argument for the distribution of source-code and associated documentation together with the system.

Prior to analysis, a \emph{preprocessing} step ensures that submissions that will be compared comply with the format expected by the detection system, and that files or parts of files which would otherwise add noise are omitted. Given the high degree of variability of submission formats, even within a given institution, this is by no means a trivial task. However, a conversion tool that can painlessly convert actual submissions into the format expected by the detection system is absent from current tools. Without it, the manual intervention required to reach a common format is likely to discourage many graders from performing automated plagiarism detection at all.

Support for result \emph{visualization} should allow a grader to judge the quality of the results of a given analysis and find those that require further inspection. Most current systems are limited to either broad, sorted-list overviews (where pairwise similarity is represented by a numerical value) or detailed, side-by-side analysis of suspect pairs of submissions. This does not allow a grader to determine whether an analysis is truly informative or too noisy to be trusted. Additionally, ranked lists cannot display or help to identify closely-related groups of submissions; and similarity is often not restricted to isolated pairs. Two notable exceptions are worth mentioning: the \textsc{MOSSCliques} \cite{popyack2003adh} system provides clique-detection within \textsc{MOSS}, and the \textsc{Same} \cite{ribler5uvd} system can generate visual representations of similarities from each submission to all others, although no higher-level overviews of test results are available.

\subsection{Types of Plagiarism and Detection Algorithms}\label{sec:state_algorithms}

A discussion on detection \emph{algorithms} and \emph{breadth} requires a brief look into the types of plagiarism-related manipulations that can be encountered, as enumerated in the following list:

\begin{description}

\item[Text replacement]\label{pt:text} Changing textual identifiers (such as variables, function names, labels), textual strings within a program, or comments. Very easy to achieve with the find/replace functionality built into most editors, although substitution of all identifiers, text and comments may require substantial time and attention to detail.

\item[Code reordering]\label{pt:order} Requires swapping the order of statements or blocks of code. Reordering can be performed internally within an instruction (e.g. \texttt{if (!A\&\&B)} transforms into \texttt{if (B\&\&!A})), within a single method or function (statement reordering) or, at a larger scope, by moving whole methods or functions within a single source file or between different source files. The cut-and-paste operation greatly facilitates this task.

\item[Code rewriting]\label{pt:rewrite} Using different syntax to express the same semantic operations as the plagiarized program, either within an instruction (e.g \texttt{i++} can be transformed into \texttt{i=i+1}) or a method/function: `if-then-else' re-engineering, loop unrolling, etc. Code reordering and re-engineering often go hand in hand, since reordering is essentially a rewriting of the sequence. Only low-level (ie.: minimal semantics) rewriting is considered here -- code rewriting at high levels of abstraction is indistinguishable from original code.

\item[Spurious code insertion/deletion]\label{pt:extra} Even in small programs, substantial parts of the code are unrelated to the core goal, and can be added or removed with few consequences. Examples include redundant error-checking or debugging statements.

\item[Source code mixing]\label{pt:mix} In a modular program, it is easy to combine code from similar implementations. Plagiarized code can be mixed with original, non-plagiarized code. It is also possible to combine code from multiple plagiarized sources, known as ``multiple plagiarism''. Mixing dilutes the traces of  manipulations that link a plagiarized submission to each of its sources.

\end{description}

Should enough different techniques and efforts be combined into a single plagiarized submission, the result can always escape detection. However, since the goal of cheating is usually to save time and effort, cheaters can be expected to stop manipulating code once the perceived risk-adjusted cost of detection drops below the cost of performing additional manipulations. In an ideal plagiarism-detection system, the cost of successful plagiarism will be greater than the cost of writing a non-plagiarized submission.

Roy and Cordy's survey on detection algorithms \cite{cordyroy2007survey} classifies existing approaches as based on matching text, token sequences, syntax trees, program dependency graphs (PDG), metrics, or a combination of these approaches. Text-based techniques achieve low recall, because they are vulnerable to all the above manipulations, but high precision, since it is rare to find the exact same textual sequences in otherwise unrelated submissions. Comparing token sequences is much more robust, since text whitespace, comments and text replacements are ignored completely, achieving higher recall at the expense of slightly more false positives. More sophisticated algorithms, based on syntax trees and dependency graphs, are capable of detecting a larger range of structural manipulations, which corresponding to larger-scope code re-orderings or rewritings. Token sequence comparison requires a lexer for each supported language. Building an abstract syntax tree for a given program requires additional language-specific support, and the use of program dependency graphs is even more demanding. Metric-based algorithms rely on extracting a fingerprint (reflecting scores on a series of metrics) from each submission, and later using only these fingerprints to find related submissions, resulting in considerable speedups. Roy and Cordy conclude that the highest recall (with reasonable precision) for simple manipulations can be achieved with token-based algorithms, and that higher-level code rewriting can only be reliably detected using program dependency graph analysis.

Given the fact that different algorithm families have very different strengths and weaknesses, it is surprising that all the systems mentioned in this section use a single similarity detection algorithm. All of them are based on different types of token sequence comparison. According to a qualitative study by Lancaster and Culwin \cite{lancaster2004csc}, two systems seemed to outperform the rest: \textsc{MOSS} and \textsc{JPlag}. However, when comparing their performance on a single corpus \cite{culwin2001scp} (there are no publicly available plagiarism-detection annotated corpora or benchmarks), neither tool significantly outperformed the other. This and other studies \cite{prechelt2002fpa,SID,liu2006gds} suggest that there is no ``silver bullet'' for plagiarism detection, and that the most critical factor regarding the degree of success (understood as reduction in plagiarism rates) achieved by a tool is the ease of use for prospective graders.


\section{Design of AC}
\label{design}

The \ac{} system seeks to address the perceived shortcomings of current systems. \ac{} is executed locally within the grader's system, and can run on a large range of platforms (the only requirement is an up-to-date Java Virtual Machine). The system and its documentation are released as open source, making \ac{} freely available to any interested parties and allowing institutions to provide their own support, in addition to that which may be available from the authors. \ac{} supports flexible submission preprocessing, can accommodate multiple similarity detection algorithms, and supports several  algorithm-independent result visualizations.

\begin{figure*}[ht]
\centering
\begin{tabular}{ccc}
\includegraphics[width=\textwidth]{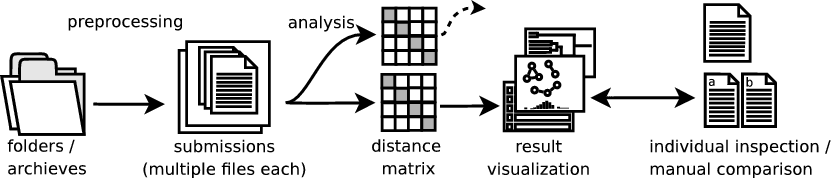} 
\end{tabular}
\caption{Flow diagram of \ac{}}
\label{fig:ac-flow-diagram}
\end{figure*}

Fig. \ref{fig:ac-flow-diagram} illustrates the conceptual plagiarism-detection pipeline used in \ac{}. Files and folders containing submissions are first preprocessed to remove irrelevant contents and accommodate them to \ac{}'s internal format. Sanitized submissions are then compared to each other, using one or more similarity algorithms. Comparison results for any given algorithm are stored in a \emph{distance matrix}, where each cell represents the ``distance'' between a single pair of submissions. These matrices can then be examined using several linked, high-level visualizations, allowing graders to explore and assess results visually. Individual inspection of particular submissions and side-by-side comparisons are available on demand. The remainder of this section describes \ac{}'s preprocessing, analysis, visualization and side-by-side comparison phases in greater detail.

\subsection{Preprocessing}

The first step when performing any type of automated plagiarism detection is to convert the set of files, folders or compressed archives that embody the submissions into a format suitable for automated processing. Apart from adapting submissions to the internal format, preprocessing should also discard irrelevant files or file segments, which would otherwise add noise to later comparisons. Performing this task manually for a large collection of heterogeneous, real-life submissions can require considerable effort. Additionally, experimenting with different sets of included and discarded files or file segments may require starting all over. Of course, preprocessing can be automated with suitable shell scripts, but such scripts are difficult to code without extensive experience, debugging them may be a time-consuming process, and they cannot be reused outside of the platform and institution for which they were originally developed. \ac{} provides an interface that automates the selection, extraction and sanitizing of submissions from their original turn-in formats into \ac{}'s internal submission input format.

Fig. \ref{fig:interface} (top) is a screenshot of the preprocessing interface, which contains two vertical panels. The leftmost vertical panel is the \emph{submission selection} panel, which allows users to select the files and folders that contain submissions. One or more views of underlying filesystems (using the familiar folder-tree metaphor) can be added to this panel, making the contents of these views selectable as ``submissions''. The rightmost vertical panel is the \emph{submission contents} panel, and allows users to view the contents of each submission that will be used in the actual comparisons. Once the user is satisfied with the submissions and their individual contents, the analysis phase can begin.

Although manual selection of submissions and pruning of submission files is possible, filters are available to automate each of these tasks. The submission selection filter allows a user to select which folders or compressed archives should be considered as ``containing submissions'' within the submission selection panel (compressed archives are extracted in-memory and treated as folders in all subsequent operations). The submission content selection filter allows individual submissions to be pruned of files which should not be included in similarity comparisons. Both filters accept nested boolean queries, which can be built and tested graphically. Testing a query highlights the files or folders that match in the corresponding pane. Atomic query terms can include regular expression matching using archive, folder, path or file contents, while composite queries aggregate subqueries using the boolean operators ``and'', ``or'' or ``nor''.

Since file names may be misleading, \ac{}'s preprocessing interface allows users to double-click on any file to see its contents, and expressions in the submission content selection filter can select files based on their actual contents (instead of relying on their names or paths).

\subsection{Analysis}

As described in section \ref{sec:state_algorithms}, many similarity-detection algorithms have been described in the literature. No single technique is superior to the rest under all circumstances. Therefore, \ac{} has been designed to be similarity algorithm-agnostic: several algorithms are available during the analysis phase, and it is easy to add new algorithms to the collection. Algorithms may also use the results of other algorithms to generate their own results. Compositions of small, specialized algorithms are easier to manage and fine-tune than larger algorithms which attempt to do everything at once.

Similarity algorithms in \ac{} are expected to write their results into a distance matrix. Each cell of this matrix represents the ``similarity distance'' between a pair of submissions. Given two submissions, $i$ and $j$, cell $\mathbf{D}_{ij}$ will be 0 if both submissions are considered identical, and near to 1 if they are very dissimilar. That is, distances are normalized to fall within the interval $[0, 1]$. Distance matrices are also expected to be symmetric, that is, the equality $\mathbf{D}_{ij}=\mathbf{D}_{ji}$ is expected to hold for every $i,j$. While distance matrices are easy to generate and manage, they do require quadratic time and space complexity, and are therefore ill-suited to very large datasets. This is not a problem for the vast majority of programming assignments.

Two main similarity detection algorithms are built into \ac{}. The \emph{normalized compression distance} algorithm uses a standard file compressor to approximate the degree of redundancy within a pair of (possibly tokenism) submissions. The \emph{token-counting} algorithm calculates a token-usage signature for each submission, and uses these for comparison. A third algorithm, the \emph{variance subtest} uses the results from a previous test and further refines its results, and is therefore an example of a composite test. Technical details of each algorithm can be found in \cite{ac-whitepaper-08}.

\subsubsection{Normalized Compression Distance}\label{zip}

A natural measure of similarity is based on the observation that two strings $a$ and $b$ are similar if the basic blocks of $a$ are in $b$ and vice versa. This is essentially how a file compressor would operate on the concatenated $ab$ sequence: the compressor would attempt to eliminate redundant information from the merged sequence; if information from $a$ is present in $b$, or vice-versa, the compressed size of $ab$, $C(ab)$, will be smaller than $C(a) + C(b)$.

\begin{equation}
\textbf{D}_{ab} = \frac{C\left(ab\right)-\min\left\{
C\left(a\right),C\left(b\right)\right\} }{\max\left\{
C\left(a\right),C\left(b\right)\right\}}
\end{equation}

where $C(x)$ denotes the length of the text $x$ compressed using some compression algorithm which asymptotically reaches the true entropy of $x$ as the length of $x$ tends to infinity.

This approach, found also in the \sys{SID} engine, has been proved to be at least as effective as those used in \sys{MOSS} and \sys{JPlag} \cite{SID}. \ac{} allows users a choice of compressors (including \sys{Zip}, \sys{GZip}, \sys{BZip2} and \sys{PPMZ}), although informal testing indicates that the default \sys{Zip} compressor is more than adequate. Tokenizing is optional; when suitable parsers are not available (Java and C/C++ are currently supported), raw text can be compared, allowing the use of \ac{} on any type of sequence at the expense of significantly lower recall ratios.

\subsubsection{Token-counting}\label{token}

This algorithm is an example of a metric-based technique, using the token distribution within tokenized submissions as the metric. For each submission $a$, a vector $\mathbf{v}_a$, containing the frequencies of each token type, is built. If two submissions share a substantial amount of code, then their token-frequency vectors are expected to be similar. Mathematically, the token-distance between submissions $a$ and $b$ is calculated as

\begin{equation}
    \textbf{D}_{ab} =
	\frac{\mathbf{v}_a}{|\mathbf{v}_a|} \cdot
	\frac{\mathbf{v}_b}{|\mathbf{v}_b|} = 
	cos(\widehat{\mathbf{v}_a \mathbf{v}_b})
\end{equation}

This is the vector-space model distance used in typical information retrieval applications, as described in
\cite{baezayates99modern}. The main advantage of this algorithm is its high speed, since comparing vectors can be implemented very efficiently. The algorithm is vulnerable against simple code rewriting, but highly resilient to code reordering (since token positions are entirely ignored).

\subsubsection{Variance Subtest}\label{variance}

If Bob copies from Alice, his submission $b$ will be much more similar to Alice's $a$ than to all other, independently developed submissions. The above algorithms are expected to assign a ``low'' value to $\mathbf{D}_{ba}$ (the distance between Bob and Alice's submissions in the distance matrix), significantly lower than all other distances in Bob's row within the matrix, $\mathbf{D}_{b}$. An interesting observation is that the exact value of $\mathbf{D}_{ba}$ is not as important as the fact that it is expected to ``stand out'' from the rest of the row. Now consider that, if Bob wishes to avoid detection after plagiarizing Alice, his best bet is to modify the source code of his submission, thereby incrementing the distance between his submission and Alice's. This will increment the distance $\mathbf{D}_{ba}$. However, unless Bob is very careful (or has many submissions to draw inspiration from), he is also likely to introduce programming constructs that nobody else has used - incrementing most distances within $\mathbf{D}_{b}$ by different degrees. Therefore, even after multiple changes to his source code, $\mathbf{D}_{ba}$ is still expected to be significantly lower than all other distances in the row (or column - the matrix is symmetrical), $\mathbf{D}_{bi}$.

The \emph{variance subtest} processes the distance matrix of a previously-run algorithm, correcting all distances to take into account their ``outlierness'' within their particular matrix row. Values that are significantly below others in their same row are decreased further (depending on their degree of ``outlierness''), to make them stand out better when represented in a histogram (see Section \ref{hist_viz}). Values that do not exhibit this a pattern are not modified at all.

\begin{figure*}[htp]
\begin{center}
\begin{tabular}{c}
\includegraphics[width=.9\textwidth]{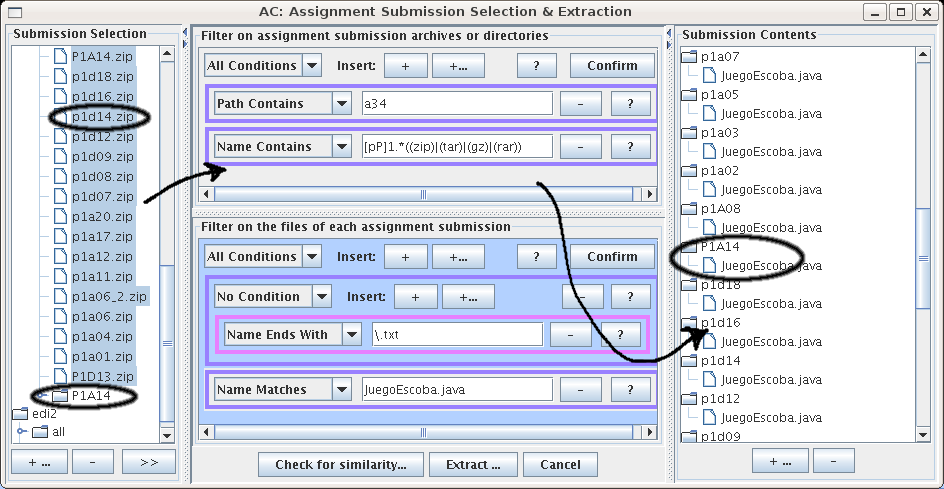} \\
\includegraphics[width=.44\textwidth]{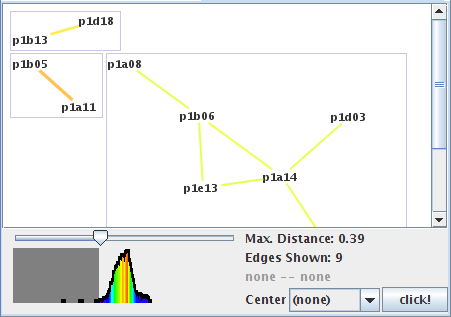}
\includegraphics[width=.44\textwidth]{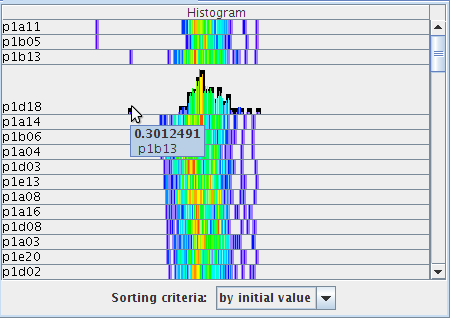} \\
\includegraphics[width=.9\textwidth]{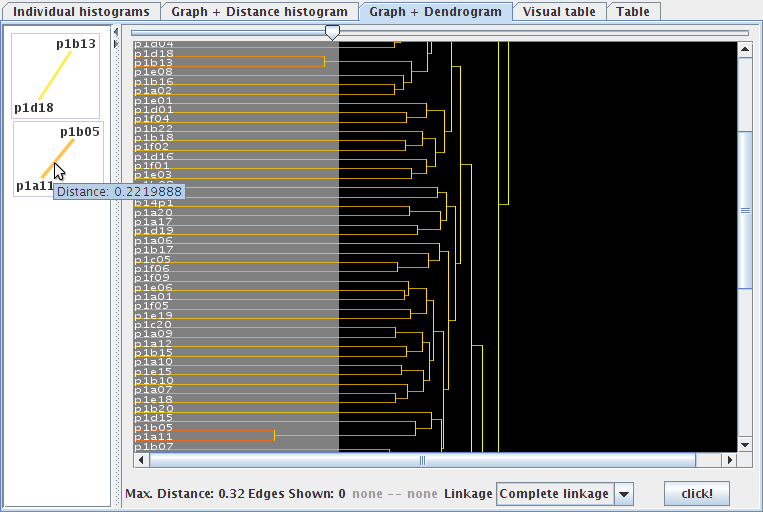} \\
\end{tabular}
\end{center}
\label{fig:interface}
\caption{Preprocessing interface (top), graph and histogram visualizations (center), and dendrogram visualization(bottom).}
\end{figure*}

\subsection{Visualization}
\label{graphicalaid}

Once an analysis is finished, users are expected to inspect the resulting distance matrix and take appropriate action if evidence of plagiarism is found. Visualization is critical to allow users to assess the meaningfulness of the results of each analysis: plagiarism detection shares many characteristics with outlier detection, where outliers are only such within the context of a broader distribution. All available visualizations can be applied to a distance matrix, regardless of how this matrix was generated. This section provides only an overview; details can be found in \cite{freire-avi08} and \cite{ac-whitepaper-08}.

Four main visualizations have been included: distance tables, graphs combined with histograms and dendrograms, and individual histograms. All visualizations provide common functionality: hovering over an area that represents a pair of submissions will display their identifiers and distances, and double-clicking the same area will display a highlighted side-by-side comparison of their contents. Double-clicking on a single submission brings up the source code for that submission.

The \emph{distance table} is the simplest of the four visualizations. It presents a sortable table of all distances, with columns for ``submission A'', ``submission B'' and ``Distance''. Sorting by increasing distances allows graders to quickly locate the greatest similarities and manually inspect the relevant submission pairs. Equivalent visualizations are common in other plagiarism-detection systems: a distance table does not provide overviews or assist graders in locating patterns.

\subsubsection{Graph Visualization}
\label{graph_viz}

Graphs are useful to visually scan for groups of submissions that are similar to each other, yet less similar to others outside this group. In the case of a submission $a$ that appears to have plagiarized from sources $b$ and $c$, a graph display can quickly answer the question of whether $b$ is also similar to $c$ or not. A screen capture of the graph visualization is presented in Fig. \ref{fig:interface} (center-left). In this graph, vertices represent submissions, and an edge is included for each similarity lower than a given threshold. Vertices without edges are omitted from the display, and for clarity, edges that are deemed redundant are also elided. Edge colors and widths are determined by the degree of similarity between their endpoints: red, thick edges denote high similarity, while green, thin edges are used to convey low similarity.

The threshold below which graphs edges are not included is set with a horizontal slider, with values ranging from 0 (leftmost position; only edges which correspond to ``exact copies'' are shown) to 1 (all edges are included). To aid the grader to select a good threshold, the slider is placed above a histogram of the frequencies of each distance in the current distance matrix. The shaded part of this histogram represents the portion of edges that are currently being considered in the graph. Graph generation and layout is delegated to \textsc{Clover} \cite{clover}, a graph visualization framework developed by one of the authors. The framework provides fast automatic layout, even for large graphs (tested to several hundreds of vertices and tens of thousands of edges).

By default, graphs views are global. However, they can also be constrained to the neighborhood of a particular submission. This allows graders to focus their attention on the selected submission, and examine its relations to surrounding submissions with less clutter than would be possible in a global view. When using a local view, the global histogram is replaced by an individual histogram which only considers distances between the selected submission and other submissions.

\subsubsection{Dendrogram Visualization}

The histogram used to set the threshold in the graph visualization can be replaced by a dendrogram, displayed in Fig. \ref{fig:interface} (bottom); the threshold slider is placed on top of the dendrogram, with the same semantics. The dendrogram is color-coded, and allows graders to group distances together according to different clustering techniques. The resulting clusters can provide additional insights into the composition of the corpus. For instance, if several instructors collaborate in a course, but provide different advice, submissions may be clustered according to the particular advice followed by each student.

\subsubsection{Histogram Visualization}
\label{hist_viz}

A third type of visualization is represented in Fig. \ref{fig:interface} (center-right). This view presents rows of ``individual histograms'', generated for each of the submissions. While the global histogram used for distance threshold selection within the graph visualization displays the frequency of \emph{all} distances found in the distance matrix, an individual histogram only displays distances for a particular row of the matrix. That is, the individual histogram for submission $A$ would be generated from $\textbf{D}_A$, the set of distances from $A$ to all other submissions.

By default, individual histograms within the histogram visualization are displayed in a condensed format where color coding, instead of bar height, is used to convey distance frequency. We have termed this abbreviated histogram a \emph{hue histogram}. The use of hue histograms allows large space savings to be achieved, and enables easy comparison of neighboring histograms. Selecting one or more histograms will expand them to the traditional height-based format; unselecting will colapse them again.

Although the histogram display may seem cryptic at first, it allows an overview of the distribution of similarities within each of the individual submissions. A key observation is that similarity ``spikes'' to the left of the similarity distribution in a given row are likely to correspond to cases of plagiarism, while a smooth individual histogram without gaps in the leftmost part of the distance distribution is an indicator that no plagiarism has taken place, since this is the expected distribution when similarity is only due to a large number of independent decisions. This is the same reasoning behind the test described in Section \ref{variance}.

\begin{figure*}[htp]
\begin{center}
\begin{tabular}{c}
\includegraphics[width=.95\textwidth]{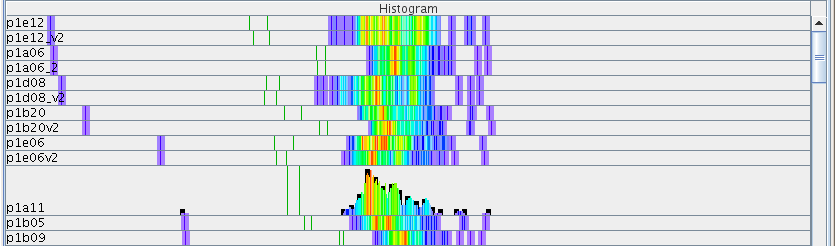} \\
\includegraphics[width=.95\textwidth]{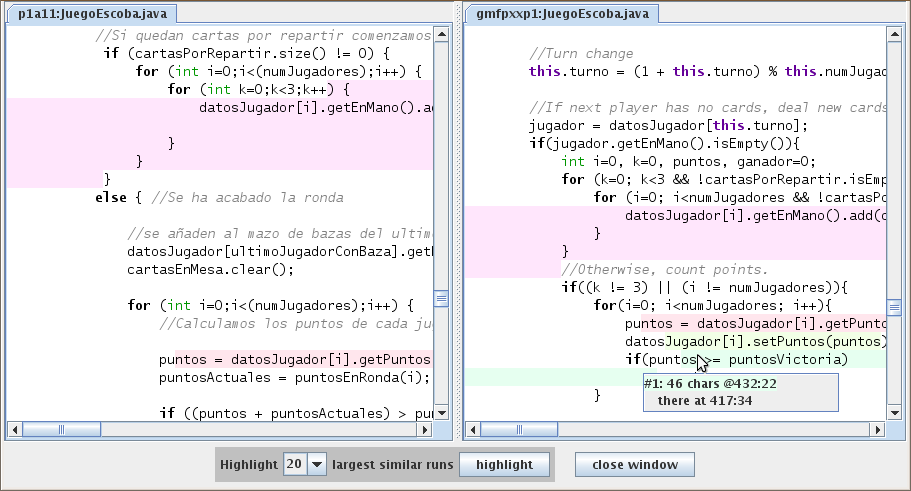} \\
\end{tabular}
\end{center}
\label{fig:hampel_and_sides}
\caption{Histograms with $0.01$ and $0.05$ Hampel identifier thresholds(top), and highlighted side-by-side source comparison (bottom).}
\end{figure*}

\subsubsection*{Threshold Recommendation using Outlier Detection}

Statistical analysis provides strong evidence that, within a plagiarism-free set of submissions, the distribution of pairwise submission distances can be adequately modelled by a normal distribution. This can be seen in the global histogram from Fig. \ref{fig:interface} (center left). From a statistical point of view, plagiarism can be considered as a generational mechanism which contaminates samples from a normal distribution with (a few) low distances. \ac{} can use the Hampel identifier, described in \cite{hampel1985bpm}, as discordancy test to extract the main tendency of the sample and characterize outliers. This test has the desireable property of not being confused by either low outliers (possible cases of plagiarism) or high outliers (extremely original submissions).

When threshold recommendation is enabled, \ac{} integrates the display of the Hampel identifier for any given histogram within the histogram itself, as a visually distinct line. The values $\alpha=0.01$ and $\alpha=0.05$ for the free parameter seem to provide good ``strict'' and ``relaxed'' thresholds. Fig. \ref{fig:hampel_and_sides} displays the positions of the $0.01$ and $0.05$ identifiers for a particular set of submissions. An in-depth analysis of outlier detection in \ac{} can be found in \cite{ac-whitepaper-08}.

\subsection{Side by Side Comparison}

Once a pair of submissions has been selected for further inspection, a side by side view of the files in each is presented to the grader. Side-by-side comparisons are found in almost all plagiarism detection systems, and usually provide a means to visually match similar blocks of code to each other.

In \ac{}, a user-selected number of duplicate fragments of code, $N$, is highlighted and color-coded in each displayed source file. Only the largest $N$ such fragments are highlighted, and the same colors are used in each side for any given fragment. Additionally, highlighted fragments have context menus that contain links to other uses of the same fragment in any of the currently compared source files. Selecting any of these links scrolls the corresponding view. A screenshot can be seen in Fig. \ref{fig:hampel_and_sides}. These context menus, unique to \ac{}, have received positive feedback by users.

\section{Evaluation}
\label{eval}

The issue of validating plagiarism detection tools has been left unattended in the literature. In personal communications \cite{Aiken2008,Joy2008}, researchers and authors of two well-known plagiarism detection tools confirmed this idea. Ideal validation requires datasets with full knowledge of which submissions have been plagiarized. Real-life datasets do not have this key feature, because of the previously mentioned problems regarding the definition and consequences of plagiarism. Indeed, the only way to obtain an accurately marked corpus is to request students to follow a specific behavior. Otherwise, carefully disguised plagiarized submissions may slip through, leading to false negatives.
Performing such an experiment would require significant time and resources, and would raise ethical concerns, since teaching students how to plagiarize is generally frowned upon.

Evaluation of \ac{} has been performed using two very different strategies. Artificially generated submissions have been used to prove that the system performs reliably in an ideal scenario, and graders at the author's institution have been requested to answer survey regarding their perceptions of plagiarism and \ac{}'s real-life effectiveness.

\subsection{Validation with Artificial Submissions}

In \cite{cebrian08}, \ac{}  is validated against artificial submissions created by means of evolutionary computing. A subset of the submissions are fully independent, and the remaining are the result of applying evolutionary operators (crossover and mutation) on top of the independent submissions. \ac{}  consistently labeled assignments with similar evolutionary origins as suspects of plagiarism. As far as we know, this approach is the only systematic attempt to validate a plagiarism detection tool. Even though automated programming and the crossover and mutation operators cannot capture many of the nuances involved in actual academic coding and plagiarism, they do provide a repeatable baseline against which plagiarism detection programs can be tested.

\subsection{Real-life Evaluation}

\ac{}  has been extensively used within the authors' institution. Since 2006, it has been used by more than 10 different lecturers in more than 15 different Computer Science courses. During this period, thousands of student submissions have been analyzed with \ac{}, and dozens of cases of plagiarism have been flagged and confirmed. Exact numbers cannot be known, because the tool has been publicly available and has been freely used by course graders. We also have reports of teachers from other institutions who have downloaded and used the tool. Students can also download the tool and its source. However, without access to a set of submissions from their peers, this should not grant them substantial advantages when trying to fool \ac{} (see Section \ref{variance}).

Use and level of satisfaction vary among graders. An informal survey developed during the last academic season gave us some insight on the actual usage patterns for \ac{}. Of the 11 respondents, 8 had experience with \ac{}, and another 3 did not use any automated tool, but expressed interest in the system. Since the number of potential respondents at this particular institution is in excess of 30, interest in plagiarism detection can be considered relatively low. This reinforces the observations from Section \ref{state} regarding the social problems of plagiarism prevention.

Of the 8 responders that had experience with \ac{}, 5 cited a large decrease in detected plagiarism between the first semester of use and further semesters, an experience that is shared by the authors of this paper. 4 respondents commented on \ac{} as being highly accurate (that is, \ac{} agrees with their final decision regarding cases of suspected plagiarism), while another 4 suspect false negatives. Indeed, one respondent commented that ``[students] have learned to plagiarize better''. Testing this hypothesis would require an experiment such as the one described at the beginning of this section. On the other hand,  ``quality'' plagiarism requires high-level understanding of what is being plagiarized, and is certainly more productive, from an educational standpoint, than simple copy and paste.

None of the respondents which had used \ac{} considered it to be confusing or difficult, and only one preferred a ranked-list based text interface to graphical representations.

\section{Conclusions and Further Work}
\label{conclusion}

Plagiarism detection is a difficult problem. The frontier between, on one side, random similarity or simple inspiration from anothers' work and plagiarism is not clear-cut, and certain cases will always require a human grader to distinguish between what is acceptable and what is not. Social issues further complicate the problem, since investing time to detect plagiarism and impose prescribed penalties is often perceived as unnecessary. Reliable manual detection is unfeasible beyond very small sets of submissions; although automated detection has been shown to be effective to identify suspects of plagiarism and flag the more complex cases, existing systems are not geared for ease of use, and support very limited types of visualizations.

This work has presented \ac{}, a plagiarism detection tool which also doubles as a framework for research into source code plagiarism detection. \ac{} directly addresses a number of concerns found in current plagiarism detection systems. The main one is that of ease of use; the human factors surrounding plagiarism make it very important to lighten the burden on graders in charge of detection, because an easy alternative is to do no systematic checking at all. \ac{} provides a partially automated graphical interface that allows submissions to be quickly converted into the system's internal format. Graphical visualizations and side-to-side comparisons allow graders to explore the general similarity distribution and quickly focus on suspect submission pairs. Statistically sound outlier detection methods are used to locate good default thresholds, futher assisting graders. These features also make \ac{} harder to fool than systems that can only provide ranked lists of pairwise submission similarity, because these lists cannot convey the context that makes any single distance truly meaningful.

Another growing concern is that of student privacy, and system availability and support. \ac{} is freely available as open source, comes with abundant documentation, and can be installed in any system with an up-to-date Java virtual machine. Graders need not worry about privacy or intellectual property concerns, since student data never leaves the grader's institution. Additionally, institutions and individuals are free to adapt and enhance \ac{} for their particular courses, without depending on the original authors.

\subsection*{Further Work}

Experimental validation can be improved upon, by gathering multiple corpora of annotated submissions. Besides the collection of anonymized and annotated classroom samples currently being prepared at the author's institution, corpora can be generated using the strategy described in \cite{cebrian08}, or even requested from student volunteers as outlined in Section \ref{eval}. The existence of a public benchmark would make the comparison of different plagiarism-detection systems much simpler, and would be of considerable aid when refining and improving these systems.

Several suggestions regarding \ac{}'s visualizations are currently being implemented. Better visualizations are expected to allow graders to gain more insight from the results of similarity tests, lowering the risk of false positives and negatives.

Another line of research is concerned with the surprisingly high accuracy of the normal distribution in outlier identification, even when confronted with different corpora and different similarity distance algorithms.

\begin{acks}
This work was supported by grant TSI 2005-08255-C07-06 of the Spanish Ministry of Education and Science. We would like to firstly thank K. Koroutchev for his seminal linux script implementation of NCD comparison.

We would also like to thank A. Su\'arez, G. Mart\'inez, J.R. Dorronsoro, M. Alfonseca, and specially P. Paalanen and S. Roberts for his technical help in the statistic analysis design of \ac{}. Additional thanks to P. Haya, L. Shafti and R. Moriyon for providing us with real submission data sets.

Finally, we would like to thank the Computer Science Department of the Universidad Aut\'onoma de Madrid for their feedback in using the tool.
\end{acks}

\bibliographystyle{acmtrans}
\bibliography{main}

\begin{received}
Received January 2008;
\end{received}

\end{document}